\documentstyle[twoside,fleqn,espcrc2,psfig]{article}

\newcommand{\AmS}{{\protect\the\textfont2
   A\kern-.1667em\lower.5ex\hbox{M}\kern-.125emS}}
\newcommand{\drawsquare}[2]{\hbox{%
\rule{#2pt}{#1pt}\hskip-#2pt%  left vertical
\rule{#1pt}{#2pt}\hskip-#1pt%  lower horizontal
\rule[#1pt]{#1pt}{#2pt}}\rule[#1pt]{#2pt}{#2pt}\hskip-#2pt%  upper horizontal
\rule{#2pt}{#1pt}}% right vertical

\newcommand{\Yfund}{\raisebox{-.5pt}{\drawsquare{6.5}{0.4}}}%  fund
\begin{document}
\title{
Renormalization 
Flow, Duality, and 
Supersymmetry Breaking in Some
$N=1$ Product-Group Theories}
\author{Erich Poppitz\address{Enrico Fermi Institute, University of Chicago,
                  5640 S. Ellis Ave, Chicago, IL 60637, USA }% 
                 \thanks{Supported by a Robert R. McCormick Fellowship and 
                   DOE contract  DE-FGP2-90ER40560. Talk given at SUSY'96, College Park, MD, May
29 - June 1, 1996.}}
\begin{abstract}
We discuss the renormalization group flow, duality, and supersymmetry breaking in  
 $N = 1$ supersymmetric $SU(N)\times SU(M)$ gauge theories.        
\end{abstract}
\maketitle
\section{Motivation}
Duality, which relates the strongly coupled behavior of one gauge theory to
the weakly coupled behavior of another, has emerged as a key idea in the understanding
of the nonperturbative dynamics of supersymmetric gauge theories.
Most of the work in this context has focused on gauge theories
with simple gauge groups.   One would like to understand theories with  
non-simple groups in more
detail: 

{\tt i.} Such an investigation will serve as a non-trivial check of simple-group
duality.  Gauging a global symmetry in two theories related by 
Seiberg duality is often a relevant perturbation,
and  the equivalence of the
resulting two theories will give further evidence for duality. 

{\tt ii.} Product groups often arise in the course of dualizing theories with
simple groups, once one goes beyond the simplest matter
representations.

{\tt iii.} Many phenomenologically interesting chiral gauge theories 
have product gauge groups.

{\tt iv.} Several classes of product-group theories exhibit dynamical supersymmetry
breaking.
\section{The Renormalization Flows}
An important question that arises in product-group theories is
whether the infrared physics
changes when one varies the ratio of the strong coupling scales 
of the two groups. A clue to the answer is given by studying the
renormalization flow in the space of the two gauge couplings. We will consider a
theory based on the gauge group $SU(N) \times SU(M)$ with a 
single (\Yfund, \Yfund) field, 
and a number of 
additional (anti)fundamentals of each group. The matter content is thus completely
specified by the number of flavors, $N_f$, of $SU(N)$ in the limit when the
$SU(M)$ gauge coupling is turned off, and the number of flavors, $M_f$, of $SU(M)$
when $SU(N)$ is turned off. 

To analyze the flows, we assume that $M, N, M_f, N_f$
 are such that in the absence of 
the other gauge coupling, each gauge theory flows to an infrared fixed point. By
turning off, say, the $SU(M)$ gauge coupling $g_M$, the $SU(N)$ gauge 
coupling $g_N$ flows to the Seiberg fixed point. Upon weakly turning on 
$g_M$, the theory flows arbitrarily close to the $g_N$ fixed point, in the 
vicinity of which  
the anomalous dimensions of 
all matter fields are approximately known (only the large 
anomalous dimensions of the   $SU(N)$ fundamentals at the fixed point,  
determined by their $R$ charge, are important in our analysis).
\begin{figure}[hbt]
\psfig{file=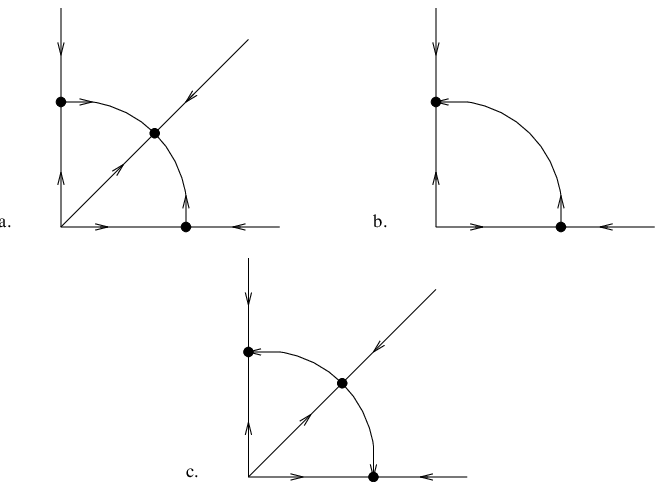}
\caption{}
\end{figure}
Using the relation between the beta function and anomalous 
dimensions,
$
\beta_{g^2} \sim - g^4 \left( 3 T(G) - \sum_R T(R)~ ( 1 - \gamma_R) \right)
$, 
we can find whether gauging $SU(M)$ is relevant at the $SU(N)$ fixed point.
Repeating the analysis in the vicinity of the $SU(M)$ fixed point, we obtain 
a set of inequalities \cite{us}, depending on $M, N, M_f, N_f$, which determine when gauging the
flavor subgroup is a relevant perturbation 
at each of the fixed points.

The three possibilities---up to interchanging the gauge 
groups---are presented on fig.1.
It is worth noting 
that only the flows from figs.1a. or 1b. occur for the allowed values of $M, N, M_f, N_f$.
The flow from fig.1c.
would imply the existence of a phase transition, as the ratio of the 
two gauge couplings is varied. There are arguments, based on holomorphy,
 against the existence of phase boundaries
in supersymmetric theories. We note, however, that the analysis of the flows above
involves the anomalous dimensions, which are not holomorphic functions in $N=1$
theories. Nevertheless the analysis is locally---in the vicinity of the two
fixed points---consistent with the absence of a phase transition.
The global nature of the flows can be studied
 in the weak coupling (Banks-Zaks) limit;
the    analysis also yields only the flows of figs.1a,b. \cite{us}.

\section{Product-Group Duality}

We next turn to studying duality in product-group theories. We analyze in detail
the case of $SU(2)_1 \times SU(2)_2$ with a single $(2,2)$ field and additional
$n$ flavors for $SU(2)_1$ and $m$ flavors for $SU(2)_2$. The
scales of the two groups are $\Lambda_1$ and $\Lambda_2$; we  call
these theories the ``$[n,m]$ models".  

Using Seiberg duality, we can  construct 
many duals of a given product-group theory.  Motivated, say, by $\Lambda_1 \gg 
\Lambda_2$, we can use Seiberg duality of $SU(2)_1$ to construct an
$SP(2 n - 4)$ dual. Weakly gauging the $SU(2)_2$ flavor symmetry, we 
obtain what we call the {\it  first dual}, 
an $SP(2 n - 4)\times SU(2)_2$ gauge theory.
Using Seiberg duality for $SU(2)_2$, we can now construct another dual theory,
the {\it second dual} $SP( 2 n - 4)\times SP(4 n + 2 m - 10)$ theory. This process
can be continued by further dualizing the first group in the second dual, etc.,
however,
the $SP$ now has an antisymmetric tensor, and its dual is still unknown. 
As an aside, we note that in the general $SU(N) \times SU(M)$ theories---where
this process can be continued by using only SQCD Seiberg duality---the chain of duals
closes, so there is only a finite number of duals that can be
constructed by using simple-group duality. Returning to the 
$[n,m]$ models, in the opposite, $\Lambda_1 \ll
\Lambda_2$, limit, we can construct other first and second duals, which have the above
groups with $m$ and $n$ interchanged. 

Our purpose is to show that all these 
duals of the original $[n,m]$ models are equivalent and independent of the 
 ratio of couplings,  the smallness of which was used to motivate the construction
of the duals.
To this end, we perform the following detailed checks \cite{us}:

1. The anomaly matching conditions are automatically satisfied, since at each step
we use Seiberg duality.

2. Consistency of duality with mass perturbations: we show that all duals,
 constructed above,
  flow under mass perturbations in a way consistent with duality, and that the
scale matching conditions for both gauge groups
 also change consistently with the  flows.

3. Consistency of duality with deformations along flat directions: we show that the 
moduli spaces of all duals are identical to that of the electric theory. 
Classical restrictions on the moduli space in the electric theory often arise 
from nonperturbative effects in the duals. In the case of product groups, 
this involves nontrivial
interplay of the nonperturbative effects in both dual gauge groups. We show
that the chiral rings of all duals are the same.

4. Mass flows to the confining phase: we show that the duals 
of the $[2,m]$ models flow to the confining phase upon adding appropriate mass
perturbations. Duality predicts that the confining superpotentials arise
through---as yet poorly understood---nonperturbative effects involving instantons in
both the broken and unbroken factors in the dual.

5. We study in detail the moduli space of the ``partially confining" models where one
of the electric gauge groups would confine if the other were turned off. 
We  show  that the dual theories reproduce the moduli space of the electric theory, 
including the quantum deformation of the  moduli space in 
the $[1,m]$ models. The  $[1,m]$ models 
exhibit another  nonperturbative effect,
specific to product-group theories: the dynamical generation of a dilaton-axion 
superfield required by Green-Schwarz anomaly cancellation. We show how duality
helps determine the dilaton field.

6. We also study the confining phase and find the exact superpotentials. The 
infrared physics in the confining phase is also independent of the ratio of couplings.

7. We note that adding  Yukawa perturbations of appropriate rank, without 
any mass terms, can also drive the theory into the confining phase. In 
the first dual, the Yukawa terms are mapped into mass terms 
that reduce the number of flavors. In the second dual, the Yukawas 
completely higgs one of the  dual 
gauge groups, and (incompletely) higgs and  reduce the number of flavors of the other 
sufficiently 
to make it confining. For smaller rank Yukawa couplings, duality predicts the 
existence of new nontrivial fixed points.

While we have performed the detailed checks for  only $M=N=2$, we expect
that our results are valid more generally, and that the obtained insight will be
useful in analyzing more  complicated product-group theories.

\section{Supersymmetry Breaking}

We show that a large class of the $SU(N) \times SU(M)$ models break supersymmetry: 
here we will only briefly 
consider the ``partially confining" models with $N_f = M$ and $M_f = N$ ($N>M$). 
The models
with $M = N -1$ are considered in  ref. \cite{us}, and the smaller-$M$  ones in \cite{usnew};
 the latter  are interesting since $SU(M)$ is not confining when $SU(N)$ is turned off.
We show that
upon adding a maximal rank Yukawa coupling, these models break supersymmetry. 
Although they possess classically flat directions, these are lifted in 
the quantum theory, and 
the models may have stable supersymmetry-breaking vacua 
(one can also add higher dimensional
operators to lift the flat directions, without 
restoring supersymmetry).

\section{Conclusions:}

\begin{enumerate}
\item{Gauging a flavor subgroup is often a relevant perturbation in a simple-group 
theory. The consistency of the flows in the electric and magnetic theories under
this perturbation provides yet another consistency check on Seiberg duality.}
\item{Our results imply that, quite generally, duals of product groups can be
constructed by using known dualities for the simple groups in the products.}
\item{The study of the renormalization flows and the detailed checks on duality
indicate that the infrared physics is independent of the ratio of the strong
coupling scales of the two gauge groups.}
\item{We showed that it is possible to flow to 
the confining phase by dimension-3 perturbations only.
This raises the possibility of constructing new phenomenologically interesting
 models of dynamical supersymmetry 
breaking.}
\item{We found new interesting nonperturbative phenomena specific to product groups,
 such as the
dynamically generated dilaton  in the ``partially confining" models.}
\item{We showed that a large subset of the  $SU(N)\times SU(M)$ 
models break supersymmetry after adding suitable dimension-3 perturbations.}
\end{enumerate}

I would like to thank my collaborators Yael Shadmi and Sandip Trivedi
for sharing their insight into the issues discussed here. Numerous helpful
discussions with David Kutasov and Lisa Randall are also gratefully 
acknowledged.

\end{document}